\documentclass[a4paper,twoside]{article}

\usepackage{epsfig}
\usepackage{subcaption}
\usepackage{calc}
\usepackage{amssymb}
\usepackage{amstext}
\usepackage{amsmath}
\usepackage{amsthm}
\usepackage{multicol}
\usepackage{pslatex}
\usepackage{apalike}
\usepackage{algorithm2e}
\usepackage[bottom]{footmisc}
\usepackage{makecell}
\usepackage{url}
\usepackage{SCITEPRESS}     

\newcommand{\wufu}{WuppieFuzz}

\begin{document}

\title{\wufu{}: Coverage-Guided, Stateful REST API Fuzzing}

\author{
\authorname{
Thomas Rooijakkers\sup{1}\orcidAuthor{0000-0001-6930-0421}, 
Anne Nijsten\sup{1}\orcidAuthor{0009-0004-5513-2948}, 
Cristian Daniele\sup{2}\orcidAuthor{0000-0001-7435-4176}, 
Erieke Weitenberg\sup{1}\orcidAuthor{0000-0001-7175-5791}, 
Ringo Groenewegen\sup{1}\orcidAuthor{0009-0005-1837-1808} and 
Arthur Melissen\sup{1}
}
\affiliation{\sup{1} The Netherlands Organisation for Applied Scientific Research (TNO), The Netherlands}
\affiliation{\sup{2} Radboud University, Nijmegen, The Netherlands}
\email{\{thomas.rooijakkers, anne.nijsten, erieke.weitenberg, ringo.groenewegen, arthur.melissen\}@tno.nl, christian.daniele@ru.nl}
}


\abstract{Many business processes currently depend on web services, often using REST APIs for communication.
REST APIs expose web service functionality through endpoints, allowing easy client interaction over the Internet.
To reduce the security risk resulting from exposed endpoints, thorough testing is desired.
Due to the generally vast number of endpoints, automated testing techniques, like fuzzing, are of interest.

This paper introduces \wufu{}, an open-source REST API fuzzer built on LibAFL, supporting white-box, grey-box and black-box fuzzing.
Using an OpenAPI specification, it can generate an initial input corpus consisting of sequences of requests.
These are mutated with REST-specific and LibAFL-provided mutators to explore different code paths in the software under test.
Guided by the measured coverage, \wufu{} then selects which request sequences to send next to reach complex states in the software under test.
In this process, it automates harness creation to reduce manual efforts often required in fuzzing.
Different kinds of reporting are provided by the fuzzer to help fixing bugs.

We evaluated our tool on the Petstore API to assess the robustness of the white-box approach and the effectiveness of different power schedules. We further monitored endpoint and code coverage over time to measure the efficacy of the approach.
}

\onecolumn \maketitle \normalsize \setcounter{footnote}{0} \vfill

\section{\uppercase{Introduction}}
Application Programming Interfaces (APIs) serve as a way of communication between (micro)services or programs that, for example, handle data or manage business processes.
In modern software architectures, APIs form a particularly important component, both for internal communication between microservices as well as receiving external input, by exposing so called API endpoints.
These may underlie a web application, making the use of APIs a preferred method, as their use is often well-defined.

Web applications provide an extensive and accessible method for organisations to present a service to their customer base.
Commonly, web APIs follow the REpresentational State Transfer (REST) architectural style \cite{fielding2000architectural}, often using the Hypertext Transfer Protocol (Secure) (HTTP(S)) \cite{martinlopez2019test}.
This way, interfaces for interaction are clearly defined, making the services easily without the need for additional software.
However, this benefit comes with additional risk.
Whilst enhancing the user experience, exposing web APIs to the outside world also introduces a new attack surface.
It is therefore crucial to detect and patch bugs and security issues as soon as possible, preferably before they end up in a production environment.
The exposed endpoints particularly pose a security risk since actors with malicious intent can easily deviate from the expected well-formed input, in an attempt to break, manipulate or disrupt the exposed services and the underlying systems.
With the vast number of exposed endpoints in this world, which grows every year, thorough manual testing for detecting issues is a nearly impossible task.
Therefore, automated solutions are desired.

Fuzzing, or fuzz testing, is a software testing technique used to discover flaws in software by feeding seemingly random inputs to the program under test (PUT).
In the last decades, the domain of fuzzing and vulnerability discovery matured with a particular focus on binary fuzzing.
State-of-the-art (coverage-guided) binary fuzzers demonstrated that the use of binary or code coverage information of the fed inputs aids in the exploration of a larger part of the SUT and, thereby, discovering more software flaws in a shorter time span \cite{wang2019be}.
As opposed to binaries, web applications often require certain sequences of input messages in order to reach a certain state, which may trigger a bug.
This, and possible inter-parameter relations, make REST API fuzzing more challenging than binary fuzzing.
Despite the domain of binary fuzzing thriving in the last few years, the research community has struggled in finding an efficient and easy to use fuzzer tailored to REST API systems.

In this paper we present \wufu{}, which is a novel coverage-guided, stateful REST API fuzzer developed with a focus on modularity, ease of use and explainability of discovered flaws for end-users.
\wufu{} supports black-box, grey-box and white-box fuzzing of REST APIs.
It is based on earlier research efforts in fuzzing as it is built on top of LibAFL, which is the first such approach in the context of REST APIs to the best of the knowledge of the authors.
Guidance based on code coverage is available for Java-based languages, Python and JavaScript with an option to extend to other languages.
To enable the use and further development of the tool, it has been made available under the Apache License Version 2.0 on a public GitHub repository
\footnote{\url{https://github.com/TNO-S3/WuppieFuzz}}.
The tool is tested for different configurations, showing that the extra work required for a white-box testing approach more quickly results in the discovery of new responses than with a black-box approach. For evaluation, the metric of response coverage is introduced, providing a more fine-grained comparison of REST API fuzzers.

\section{\uppercase{Background}}
In this section, we present the background and terminology used in the paper.
Also, we introduce LibAFL, the framework we used to build \wufu{} as well as previous work on REST API fuzzing.

\subsection{(Binary) Fuzzing}
Fuzzing is a software testing technique used to find vulnerabilities in software~\cite{ossfuzz2023ossfuzz}.
When fuzzing a system, malformed inputs are sent to the SUT in an attempt to trigger unexpected system behaviour. The fuzzers then give the analyst information about the exact input that triggered the bug.
It is worth it to mention that the fuzzers usually employ complex strategies to tailor the generation of input toward the most interesting ones, as completely random messages will very likely bring poor fuzzing performance.
According to the input generation, the fuzzers are divided into mutation and generation based.

\par{\textbf{Mutation-based fuzzing.}}
Initially, fuzzers used a pure random, brute-force strategy to generate input (data) to feed to the SUT.
As already mentioned, generating totally random input is very likely to achieve poor performance.
While a random component is still part of the fuzzing process, smarter fuzzers do not generate inputs from scratch.
Instead, they generate input data by performing mutations on an initial set of inputs, often called seeds.
A seed is a (well-formed) input that serves as a starting point for mutational input generation.
The initial set can consist of one or more seeds, and during the fuzzing campaign, the set can be expanded, according to the findings of the fuzzer.
This collection of seeds is also known as a seed corpus.
Different fuzzers perform different mutations over the seed files.
Examples of mutations are flipping bits and recombining (parts of) seeds, also known as splicing.

\par{\textbf{Generation-based fuzzing.}}
Instead of a set of seeds, generation-based fuzzers use some knowledge of the data structure on the inputs expected by the SUT.

This knowledge (often expressed in terms of grammar and state model) can be inferred from the documentation of the SUT, or from the analysis of the network traffic. Based on this grammar, new inputs are generated.
The new inputs are most likely to pass the initial parsing checks, allowing the fuzzer to reach deeper portions of the software. However, this approach requires in-depth knowledge of the protocol being used which might limit the scalability of the approach.

\begin{figure}[ht]
	\centering
	\includegraphics[width=0.8\linewidth]{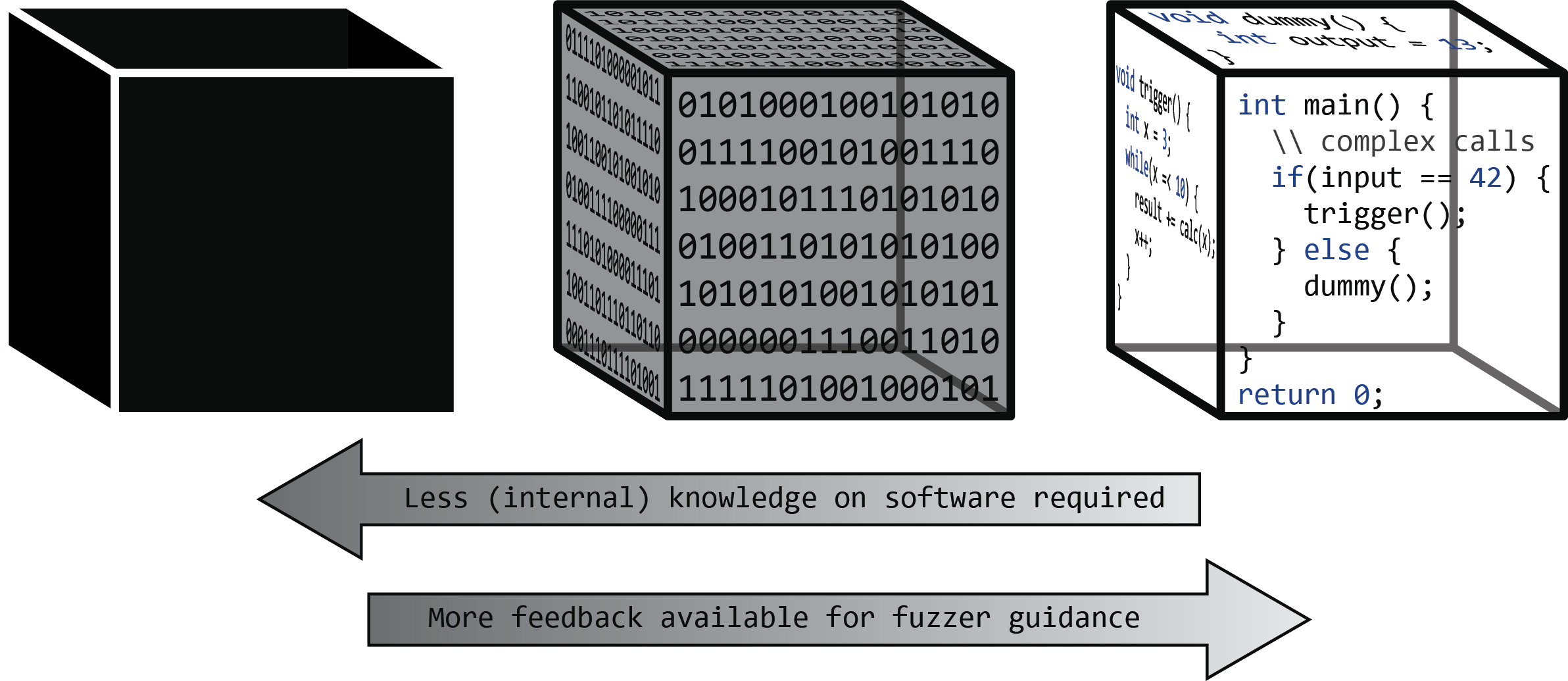}
	\caption{Visualisation of the classification of possible settings based on the amount of (internal) knowledge of the SUT.}
	\label{fig:setting}
\end{figure}

An orthogonal classification of fuzzers has been made according to their knowledge of the SUT, as depicted in Figure~\ref{fig:setting}.

\par{\textbf{Black-box fuzzing.}} As the name suggests, in black-box fuzzing the SUT is handled as a black-box, which means that the fuzzer feeds input data to the SUT and the SUT returns an output which is then retrieved by the fuzzer.
This corresponds to the nature of many web applications that are only available as a black box to the user.
In this setting only information that can be measured from the outside can be taken into account, e.g. response time and responses.

\par{\textbf{White-box fuzzing.}}
The white-box terminology is linked to transparency of the software.
In this setting, there is a lot of knowledge that can be used as feedback to guide the fuzzer.
In particular, we have full vision on the internals of the software, for instance, by having access to the source code of a compiled software binary.
This additional information can be used to guide the fuzzer and also helps in the search of what exactly caused unexpected behaviour, and is thus interesting when source code is available, e.g. for developers or users of open-source code.

\par{\textbf{Grey-box fuzzing.}}
In between we have grey-box fuzzing, which is a setting where there is no full transparency, nor black-box opacity.
An example of grey-box fuzzing would be a setting where there is no access to the source code, but we have full access to the compiled software.
This access enables fuzzers, for instance, to track the control flow of the SUT through binary instrumentation, thereby generating feedback on what parts of the binary were visited by what input. 

\subsection{Coverage-Guided fuzzing}
When information about what lines of code have been triggered by some input, i.e. code coverage information, is used to make smart(er) choices on what random inputs to try, e.g. how to mutate input data, this is called coverage-guided fuzzing.
Information on code coverage can be found in Section \ref{sec:codecoverage}.
Because many mutations can be constructed from a single input, this guidance helps prioritising mutations that create more code coverage.
This way, bugs that are located deeper in the software program can be found.

The instrumentation needed for coverage-guided fuzzing comes at a trade-off.
The use of code coverage information can improve the fuzzing results, however, it also introduces overhead during setup and runtime.
On the contrary, black-box fuzzing no additional instrumentation needs to be added to the SUT and can therefore be performed more easily and without any slowdown during testing.
This trade-off should be considered for every fuzzing use case.

For example, when there is only time for a quick fuzzing campaign black-box fuzzing will be more interesting, while in a CI/CD environment white-box coverage-guided fuzzing may be of more interest.

\subsection{Challenges}
Multiple challenges may arise with the use of different fuzzers. The following is a non exhaustive list of some of these challenges.
Firstly, creation of the harness, which manages the interface between the SUT and the fuzzer for giving input and receiving feedback, is not always straightforward.
Secondly, reaching of states that exhibit bugs that are reachable only after a sequence of inputs has been given to the SUT (stateful bugs). In a REST API setting, the grammar required needs to keep into account multiple factors, like message structure, message order and parameter relations, which increases the complexity.
Thirdly, interpretation of the results given by the fuzzer, as randomised input may give very unexpected results.
Lastly, configuration of the fuzzer in order to get it running quickly when convenient, as well as receive bug reports in a way similar to other test cases for the SUT.
\wufu{} addresses these challenges, as explained in Section \ref{sec:wufu} and summarised in the concluding Section \ref{sec:conclusion}.

\subsection{Code Coverage}
\label{sec:codecoverage}
Code coverage is a well-known metric used in software testing, and high code coverage is therefore a desirable result, as one cannot say anything about the correctness of parts of the code that have not been covered \cite{hemmeti2015how}.
However, it should be noted that a hundred percent code coverage does not necessarily mean the code does not contain vulnerabilities, nor that it is exhaustively tested \cite{dijkstra1970notes}.
There exist many different instrumentation tool kits for measuring and reporting code coverage.
Examples are JaCoCo for Java (byte code) \cite{eclemma2024jacoco} and coverage.py for Python \cite{batchelder2024coveragepy}.
When source code is available, it can be instrumented during compilation to measure e.g. line coverage or statement coverage.
When only a binary is available, the byte code can be instrumented dynamically during runtime or statically before runtime.
While dynamic instrumentation allows for more accurate analysis, it also causes a significant performance overhead \cite{lou2020study}.

\subsection{LibAFL}
American Fuzzy Lop (AFL) is an open source, coverage-guided fuzzer which was considered the state-of-the-art for a while.
Many contributions to improve this fuzzer were made in parallel.
LibAFL was presented as an effort that combined these contributions into a framework for building modular and reusable fuzzers~\cite{fioraldi2022libafl}.
This framework can be used to develop a custom fuzzer, which can easily be extended with new developments.

\subsection{REST}
REST is one of the most popular architectural styles for describing machine-to-machine interfaces and is often used in the development of web applications in a client-server model \cite{martinlopez2019test}.
REST uses a request-response model, generally formatted in HTTP.
HTTP methods, such as \texttt{POST}, \texttt{GET}, \texttt{PUT}, and \texttt{DELETE} are used to manipulate resources at the API endpoint by creating, reading, updating, or deleting those, respectively.
These four basic operations for persistent (data) storage are also referred to as Create Read Update and Delete (CRUD).
The request's payload (and its response) is generally formatted as plaintext, XML, or JSON.

\subsection{OpenAPI}
The OpenAPI specification, formerly known as Swagger specification, is a widely used standard for describing the request-response model of a REST API \cite{dimartino2018a,openapi2024openapi}.
The OpenAPI specification defines supported endpoints (paths), supported (HTTP) methods, expected responses, expected HTTP response status codes and parameter and request body formats and examples.

\section{\uppercase{Related work}}
While fuzzing efforts for binaries lead to a plethora of binary fuzzers being available over the years, developments in fuzzing of web applications have gained popularity more recently.
This section highlights some API fuzzers, yet this is by no means extensive nor covers all methodologies used by API fuzzers.
For more detailed, recent surveys, the reader is referred to \cite{dharmaadi2025fuzzing,golmohammadi2023}.
API fuzzers are available both freely and commercially, oftentimes published open source.
Some fuzzers focus on detecting specific vulnerabilities, like those described in the OWASP API Security Top 10, while others are more general purpose, looking for crashes or unexpected behaviour.
Most literature focuses on black-box API fuzzing.

A popular example of such a fuzzer is RESTler, a stateful black-box REST API fuzzer developed by Microsoft, that uses OpenAPI specifications to generate a fuzzing grammar \cite{atlidakis2019restler}.
It detects bugs by checking server errors and the correctness of responses against the specification.

Recently, EvoMaster was claimed to be the only white-box coverage-guided API fuzzer available, while also offering a grey-box and black-box mode, despite existing for several years and publishing promising results \cite{arcuri2024advanced,arcuri2017restful,zhang2023open}.
It currently only supports white-box testing for APIs compiled to JVM, like Java and Kotlin, and previously had unstable support for C\# and JavaScript/TypeScript~\cite{evomaster2024evomaster}.

Grey-box web API fuzzers like BackREST are also available \cite{gauthier2022experience}.
Another early example of grey-box fuzzing aimed at detecting Reflective Cross Site Scripting bugs is webFuzz~\cite{rooy2020an}.

Code Intelligence claims to offer coverage-guided REST API fuzzing as part of their commercially available CI Fuzz platform with a focus on CI/CD integration~\cite{resch2024continuous}.

Lastly, GitLab also offers CI/CD focused API fuzzing and coverage-guided fuzzing services as part of their paid plans, but not coverage-guided API fuzzing~\cite{gitlab2024gitlab}.

\section{\uppercase{\wufu{}}}
\label{sec:wufu}
In the following section the overall architecture of the coverage-guided, stateful REST API fuzzer \wufu{} is presented, as well as its different components and some implementation considerations.

\subsection{Architecture Overview}
\begin{figure}[ht]
	\centering
	\includegraphics[width=\linewidth]{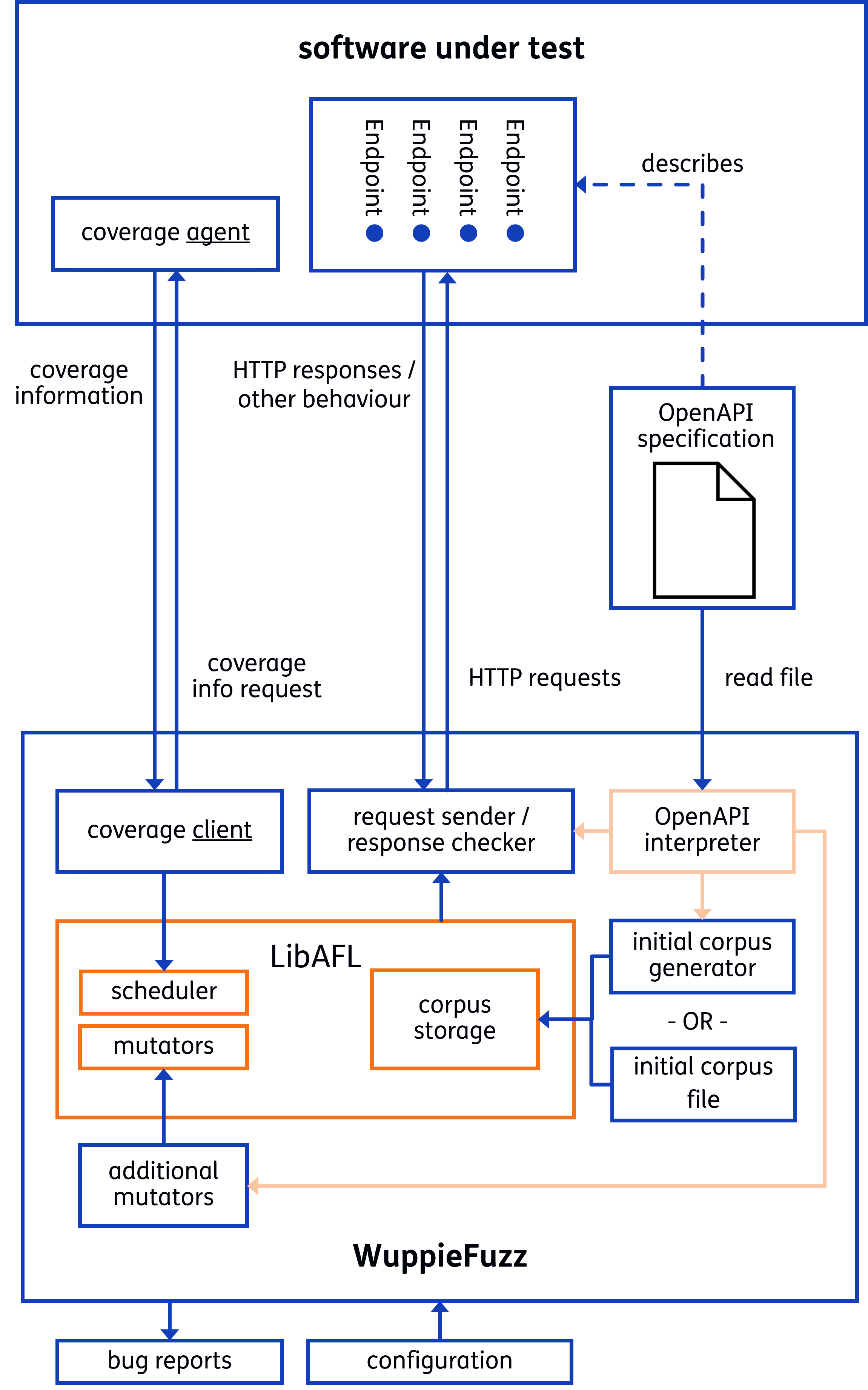}
	\caption{Visualisation of the interaction between the SUT and \wufu{}, as well as the different components of the fuzzer. The LibAFL library is shown in orange. The OpenAPI interpreter and the different places where its results are used are shown in light orange.}
	\label{fig:architecture}
\end{figure}

Communication between \wufu{} and the SUT occurs through two main channels.
Firstly, the HTTP requests and responses, and secondly, the coverage information requests and responses.
In this section, these interactions between the fuzzer and the SUT will be explained.
The overall architecture is depicted in Figure~\ref{fig:architecture}.

\subsubsection{OpenAPI Specification}
\label{sec:openapispec}
An OpenAPI specification that describes the SUT must be available, as it is automatically interpreted by \wufu{}.
This information is used in several places, as can be seen in light orange in Figure~\ref{fig:architecture}.

Firstly, the response checker in the fuzzer can use the specification to check whether there are unexpected responses.
In case no expected responses are specified, the response checker can also be configured to flag only HTTP status codes that indicate a server error.

Secondly, the initial seed corpus is either generated by the fuzzer based on the specification, or provided as an input file by the user of the fuzzer.

Thirdly, several mutators also use the specification to make mutations on the request sequence that is sent to the SUT, e.g. to create and break relations between request parameters and responses.

\subsubsection{Harness}
\wufu{} includes an HTTP client which sends out HTTP requests to the endpoints of the SUT and receives HTTP responses.
Other responses, like timeouts and connection closure signals are also received.
These responses are verified for conformance to the API specification by the response checker.
Together, the HTTP client and response checker form the harness.
As described in the previous paragraph, this harness creation is an automated process.
Clearly, this requires the SUT to expose a REST API.

When a sequence of requests (seed) is selected, the structure of the seed and/or individual requests are mutated to generate test cases that are fed to the SUT.

\subsubsection{Coverage Monitor}
The prioritisation is determined by the scheduler based on coverage (code coverage and/or endpoint coverage) information, which the fuzzer's coverage client receives from the coverage agent in the SUT.
Here, coverage of an endpoint is determined by the combination of the path, method and returned status code.
\wufu{} supports coverage agents for several programming languages, like Java, Python and JavaScript.
This is done in a modular way so it can be easily extended with coverage agents for other languages.
In case there is no coverage agent available or no extension has been made for an unsupported agent, black-box testing is still possible by only measuring endpoint coverage.
This will however provide less detailed feedback than line coverage provided by a coverage agent, and thus may result in a less effective fuzzing campaign.

\subsubsection{Reporting and Configuration} \wufu{} can be configured using a configuration file or command line arguments.
Among others, these settings allow for selection of a coverage agent and authentication mode.
Flags can be passed to the fuzzer to enable the generation of several kinds of reports for each fuzzing campaign, containing coverage information and details about the requests and received responses.

\subsection{Fuzzer Components}
In this section, the internal components of \wufu{} as introduced before will be elaborated on. Further, the general workflow when using the fuzzer will be explained.

\subsubsection{Seed Corpus Generator}
\label{sec:scgen}

Seed inputs are often (examples of) correct inputs for the SUT, as a starting point for the fuzzer so it quickly finds the happy code paths in the SUT.
By itself, it would be very difficult, or even nearly impossible in the case of complicated REST APIs, for the fuzzer to find such a path.
Constructing these seeds by hand is possible, but time consuming.
Therefore, \wufu{} uses the OpenAPI specification of the SUT to construct seeds in an automated fashion.

\begin{figure}[ht]
	\centering
	\includegraphics[width=\linewidth]{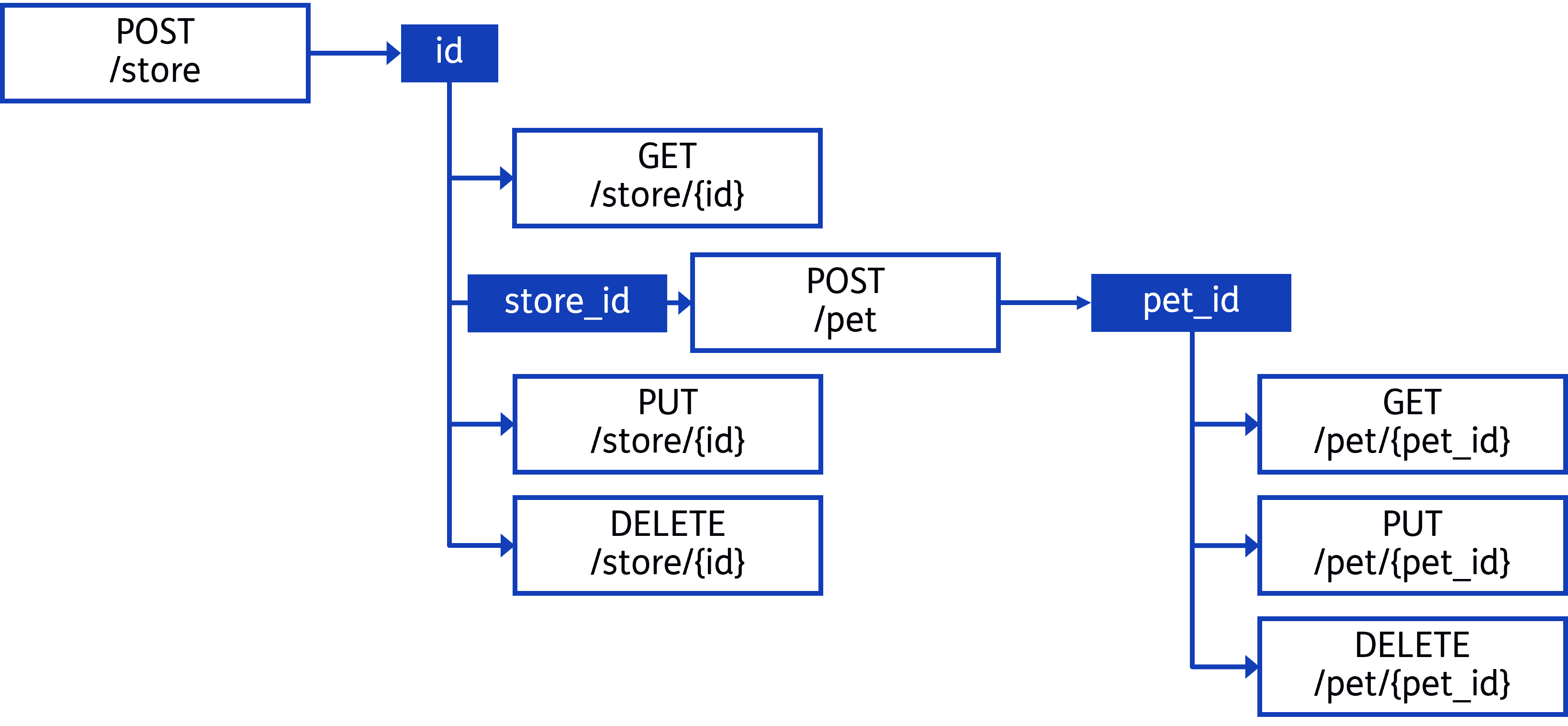}
	\caption{Example graph, showing the relations between API endpoints through the use of resources (dark boxes) created by other endpoints.}
	\label{fig:corpusgen}
\end{figure}

First, the seed corpus generator tries to understand the connections between the API endpoints of the SUT by generating one or more graphs.
For each resource controlled by an endpoint, it finds the other endpoints that may make use of this resource, and sorts them in CRUD order.
The possible use of the resource at different endpoints can be indicated by using the same or similar names for parameters.
Depending on the path in a request, prepending or postpending elements to a parameter name may provide context to the parameter.
Due to so-called stemming steps taken by the generator, which processes words into different categories, use of similar words like ``store'' and ``stores'' can be related to the same context.
To illustrate this, an example can be found in Figure~\ref{fig:corpusgen}.
In this example, a pet store can be created with \texttt{POST /store}, resulting in a response that includes an \texttt{id} resource.
This \texttt{id} can then be used as a path parameter to read, update and delete information about the pet store with \texttt{GET}, \texttt{PUT} and \texttt{DELETE}, respectively.
In the graph, the requests are sorted in CRUD order.
Furthermore, the \texttt{POST /pet} request, that requires a \texttt{store\_id}, can be used to create an entry for a pet in the store.
This then returns a \texttt{pet\_id}, which is in turn used for its respective read, update and delete operations.
Because the \texttt{store\_id} parameter name needed for \texttt{POST /pet} is similar to the \texttt{id} returned in the response to \texttt{POST /store}, this relation between \texttt{id} and \texttt{store\_id} is added to the graph.
If there would be another parameter used somewhere in the SUT called \texttt{stores\_id}, it will be related to \texttt{store\_id} in the graph, as the stemming text processing will put them in the same category.

With this interpretation of the relations, HTTP requests with a sensible order can be generated.
When a response to a certain request contains a value in a field that corresponds to a field in a subsequent request, this value can then be placed in this corresponding field.
Other fields are filled with example values taken from the OpenAPI specification when available, generated based on a specified pattern, or default values for the data types that are specified for them otherwise.

This way, sequences of HTTP requests that can be accepted by the SUT can be generated.
When the OpenAPI specification covers all functionality of the API, this results in a corpus that covers all available API operations. The user may save this corpus to make manual adjustments, as discussed in Section \ref{sec:reporting}.

\subsubsection{Mutators}
Mutators are used to make changes to seeds in the corpus, which each are sequences of API requests.
By initiating a fuzzing campaign with a corpus containing valid inputs, provided by the corpus generator or by the user, subsequent mutations of these inputs will have a greater chance of being valid enough to traverse a path within the SUT and thus increase code coverage.
For the same reason, mutations are not directly applied on parameter values that were obtained from responses, only indirectly by e.g. the BreakLinkMutator as described in Table~\ref{tab:wufumut}.

A variety of mutators are employed, each of which makes their own kind of changes, to create a large array of mutations.
\wufu{} uses both mutators provided by LibAFL and additional custom HTTP-specific mutators.
The LibAFL mutators are primarily used for making low-level alterations to request parameters, whereas the custom mutators mostly change the structure of the sequence of requests, as this functionality is not provided by LibAFL mutators.
Examples of LibAFL and custom mutators are inserting bytes and swapping requests, respectively. More details on these mutators can be found in Appendix~\ref{app:mutators}.

From the available sequences of requests, the fuzzer chooses an interesting sequence as described in the following sections.
It then applies randomly chosen mutations from the 31 available mutators, resulting in a request sequence that will be sent to the SUT.

\subsubsection{Scheduling}
There are a variety of metrics that can be used to prioritise one sequence of requests generated by making some mutations over another.
\wufu{} uses the FAST power scheduling algorithm provided by LibAFL, which provides a measurement on how ``interesting'' an input is \cite{fioraldi2022libafl}.
Examples of ``interesting'' inputs are ones that visit paths that have not been visited yet or that are not visited often. Additionally, inputs that execute more quickly are more ``interesting'' in the FAST power schedule.
Details on power schedules, which was originally designed for binary programs, are discussed in \cite{bohme2016coverage}.
Among the metrics used by the power scheduler to prioritise the inputs are coverage information, as well as the number of requests in a request sequence.

\subsubsection{Coverage Information}
It is desirable that the fuzzer has a high code coverage, testing as many of the paths in the code as possible.
\wufu{} tracks what inputs lead to what code coverage in the SUT.
Inputs that lead to new code coverage are prioritised for mutation.

To obtain line coverage information, the SUT runs with a coverage agent, i.e. the SUT is instrumented.
The coverage agent is exposed as a service on a separate port, allowing the fuzzer's coverage client to retrieve the coverage information and reset the tracked coverage after each sequence of requests.
Note that usage of a coverage agent is not an option for all SUTs, e.g. when their language is not supported or when it is hard to incorporate a coverage agent.
In those cases, only endpoint coverage is used.

The coverage information is translated into a bit vector.
For line coverage, this is a vector with the number of lines in the code as length, where ones indicate covered lines.
For endpoint coverage, the vector contains a bit for each expected HTTP status code for each API method and path, and is extended when unexpected status codes are encountered.
Here, each vector contains a single bit with value one, indicating the response for a certain endpoint that was received.

\subsubsection{Workflow}
Here, we sketch the order of events occuring in the usage of \wufu{} based on the previously described components.
First, with \wufu{} configured as desired and and the SUT, along with an OpenAPI specificaton, in place, a seed corpus is generated.
Following is an iterative process.
A sequence of requests from the corpus is mutated by LibAFL and/or custom mutators.
Using the LibAFL scheduling functionality, the next sequence of requests to be sent to the SUT is chosen.
When the sequences in the request are sent, responses are received and checked against the OpenAPI specification.
Further, related coverage information is requested from the SUT and received in case of white-box fuzzing.
This coverage information can then be used by the scheduler.
Lastly, reporting information can be obtained from \wufu{}.

\subsection{Implementation}
The described fuzzer has been implemented in Rust, because of its performance and to remain consistent with LibAFL, which was also developed in Rust. 
The source code has been made available on a public GitHub repository to be open for use and contribution. 

For usability purposes, several options on reporting and configuration of the tool are available.
Here, the reporting options will be elaborated upon.
The full list of configuration options is available in the fuzzer documentation.

\subsubsection{Reporting}
\label{sec:reporting}
The fuzzer can generate different types of reports, with a twofold purpose.
Firstly, it gives insights in the performance of the tool and its configuration.
Secondly, the reporting can be used to study and solve issues found by the fuzzer.

The API dependency graph as well as the initial corpus created by the seed corpus generator, as described in Section \ref{sec:scgen}, can both be reported in Markdown/Mermaid format.

\begin{figure}[ht]
	\centering
	\includegraphics[width=\linewidth]{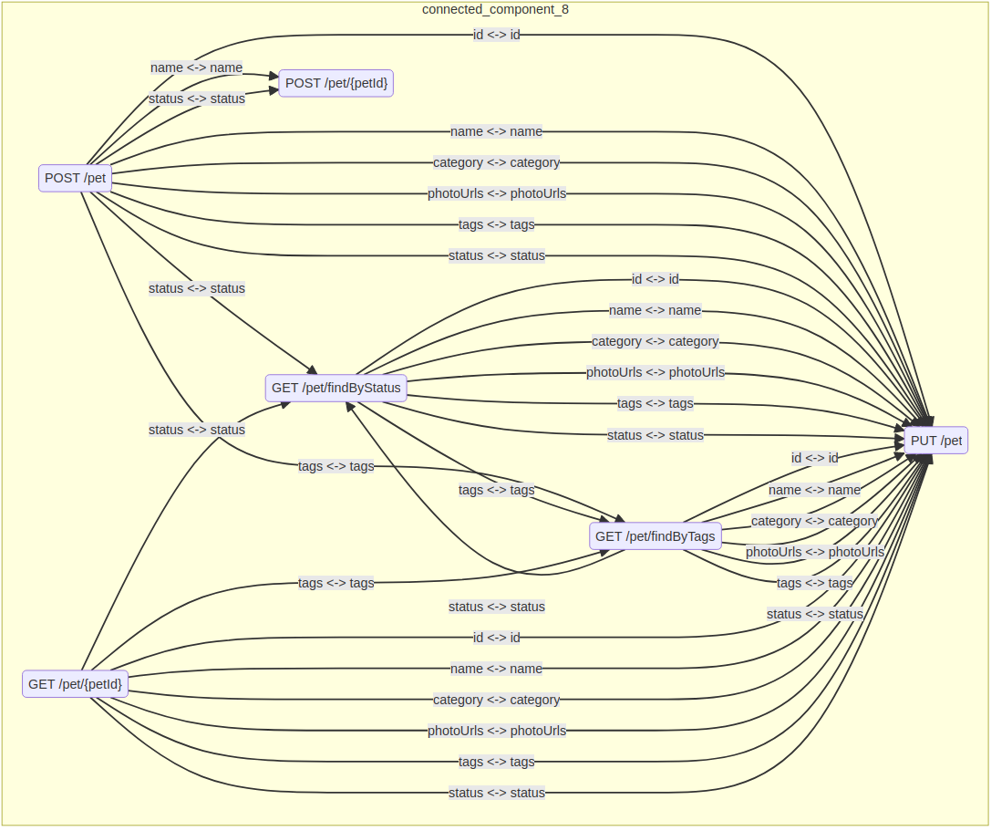}
	\caption{Example API endpoint dependency graph as reported by \wufu{}.}
	\label{fig:dependencygraph}
\end{figure}

The API dependency graph can be generated as a directed graph, where nodes are combinations of available endpoints and HTTP methods.
An example of such a graph can be found in Figure \ref{fig:dependencygraph}.
The edges are directed from one node \textit{A} to another node \textit{B} if information from the response of a request with method and endpoint as specified in \textit{A} is used in the request with method and endpoint as specified in \textit{B}.
A label on the edge specifies which value has been detected by the tool that can be reused in the next request from the response to the previous request.
There may be multiple edges between each pair of nodes to indicate reuse of multiple values.
The information in the graph can be used to check whether the fuzzer was able to deduce all interesting relationships between resources from the OpenAPI specification.

The initial corpus can also be reported as a set of graphs, depicting a sequence of requests for the SUT.
This can be used to check whether the initial corpus indeed contains sensible values.
If this is not the case, or when the user wants to apply other changes, this can be updated in the corpus manually.

A report on the endpoint coverage of the fuzzer on the SUT can be generated, showing endpoints and their related status codes, either as given in the OpenAPI specification, or others that were found during fuzzing.
A check mark, cross and warning sign indicate that a specified status code has been triggered, a specified status code has not been triggered and that an unspecified status code has been triggered, respectively.
One can click on a triggered status code to see an example request that triggered it.

\begin{figure}[ht]
	\centering
	\includegraphics[width=\linewidth]{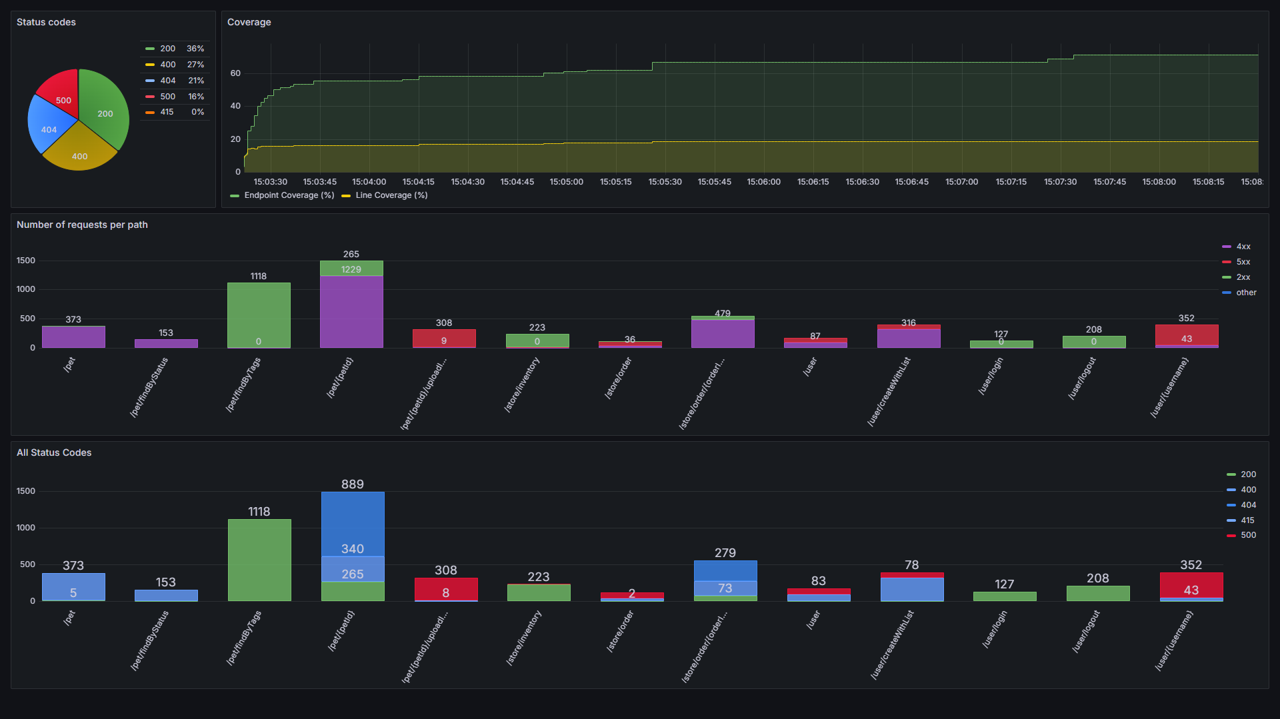}
	\caption{Example of information displayed on the dashboard.}
	\label{fig:dashboard}
\end{figure}

There is no reset of the SUT in between input sequences, so it is possible that found bugs result from a state that was generated by earlier inputs.
Storing information like responses and inferred values together with the behaviour of the SUT may help to examine these bugs.
To help make this information insightful, as the fuzzer generates a vast number of requests, a SQLite database and Grafana dashboard are provided.
This shows a pie chart of received status codes, the endpoint and line coverage over time, as well as a bar chart depicting the number of requests and received status codes for each endpoint, as can be seen in Figure~\ref{fig:dashboard}.
It also shows where to find more information on these requests.

Lastly, the fuzzer also has the option to generate a code coverage report when the source code of the SUT is available and the language is supported by \wufu{}, containing information on which lines of code have been executed.

Together, this information can be used by e.g. developers or pentesters to find out more about unexpected behaviour caused by the fuzzer.
This can then either be fixed or investigated for further exploitation.

\section{\uppercase{Experimental evaluation}}
\begin{table*}
    \centering
    \caption{Response coverage of black and white box approach of the Pet Store REST API system after 10 minutes of fuzzing.}
        \begin{tabular}{|c|c|c|c|c|}
            \hline
          \textbf{Scheduler} & \multicolumn{2}{c|}{\textbf{Black-box}} &\multicolumn{2}{c|}{\textbf{White-box}}\\
           & \textbf{Coverage} & \textbf{Requests sent}& \textbf{Coverage} & \textbf{Requests sent}\\
          \hline
          Fast & 19/58 &149642& 19/58 & 13250\\
          \hline
          Explore & 19/58 &132011&18/58&15454 \\
          \hline
          Lin &19/58&132640&18/58&12967\\
          \hline
          Exploit &19/58&134129&19/58&17835\\
          \hline
          Quad & 18/58 &127761& 20/58&14658\\
          \hline
          Coe & 20/58&126680&17/58&13911\\
           \hline
        \end{tabular}
    \label{tab:schedulers}
\end{table*}

We tested \wufu{} on the Petstore\footnote{https://github.com/swagger-api/swagger-petstore} REST API system. We fuzzed the system using different schedulers, with a white-box and black-box approach in order to evaluate difference in functionality. In the evaluation, different types of coverage are measured.
\textit{Endpoint coverage} shows how many endpoints were tested and returned a correct response code (2xx).
Although this metric is widely used in the REST API fuzzing community, it lacks information on how many responses were triggered for each endpoint.
Besides the endpoint coverage, \wufu{} monitors the \textit{response coverage}, which indicates the number of different responses received for the different endpoints of a SUT, to give a more fine-grained benchmark.
Lastly, line coverage over time was measured.

\subsection{Schedulers}
In order to evaluate the impact of different schedulers, we run \wufu{} over the Petstore REST API program with the six schedulers offered by LibAFL. As shown in Table~\ref{tab:schedulers}, in the black-box mode, all schedulers achieved comparable endpoint coverage (ranging between 18/58 and 20/58). This suggest that no single approach outperforms the others. However, the \textit{Fast} scheduler generated the highest number of requests ($\sim150k$), while Coe achieved the same or higher coverage with roughly 15\% fewer requests. This suggests that Coe more intelligently focuses on the most interesting messages to fuzz.

In the white-box mode, coverage values are again close, between 17/58 and 20/58, but the required number of requests is different. In fact, the Exploit scheduler --- which emphasizes mutations of already promising inputs --- reached similar coverage to Fast but with a significantly higher number of requests ($\sim17.8k$). On the other hand, Quad achieved the highest coverage (20/58) with $\sim14.6k$ messages, demonstrating a good balance between exploration and exploitation. The Coe scheduler underperformed slightly in this mode (17/58), indicating that its coverage-guided heuristics may be less effective when instrumentation feedback (e.g., from JaCoCo) is available.

\subsection{White- vs black-box approach}

\begin{figure}[ht]
	\centering
	\includegraphics[width=\linewidth]{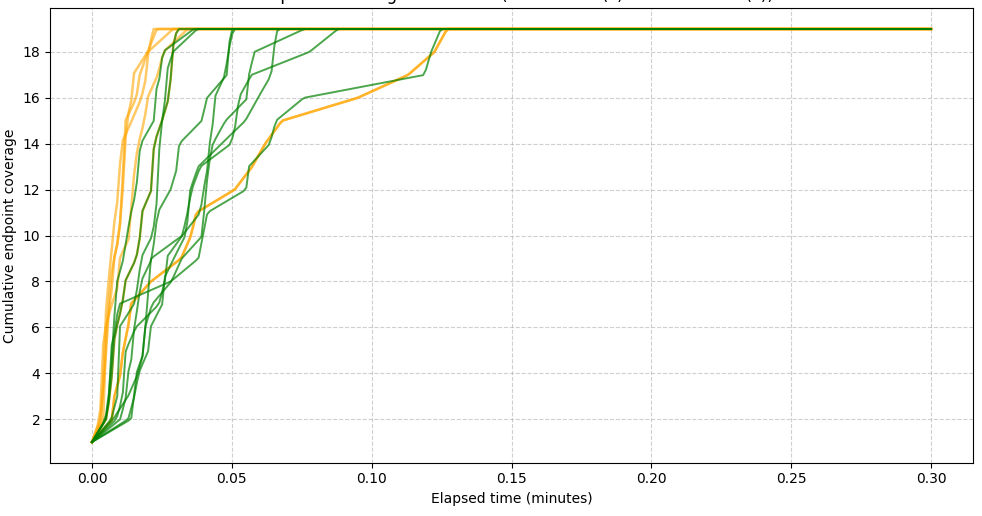}
	\caption{Response coverage of black-box (in green) and white-box (in orange) approach.}
	\label{fig:comparison_endpoint_coverage}
\end{figure}

Figure~\ref{fig:comparison_endpoint_coverage} shows the cumulative endpoint coverage over time for ten fuzzing runs on the Petstore target.
Each curve represents a different fuzzing session. Orange lines represent the white-box runs --- guided by code instrumentation --- and the green lines the black-box ones.

The white-box setting achieves a slightly faster endpoint discovery and reaches a plateau at 19 endpoint responses. All the different measured responses are triggered within the first 10 seconds.
The black-box approach shows larger difference across the runs. While some runs reach the coverage almost as quickly as the white-box runs, others seem more sensitive to the initial seed selection or to the random exploration order.

The coverage achieved by both approaches is the same, implying that --- in this specific SUT ---, the white-box approach slightly accelerates the discovery but does not increase the absolute endpoint coverage. 

\begin{figure}[ht]
	\centering
	\includegraphics[width=\linewidth]{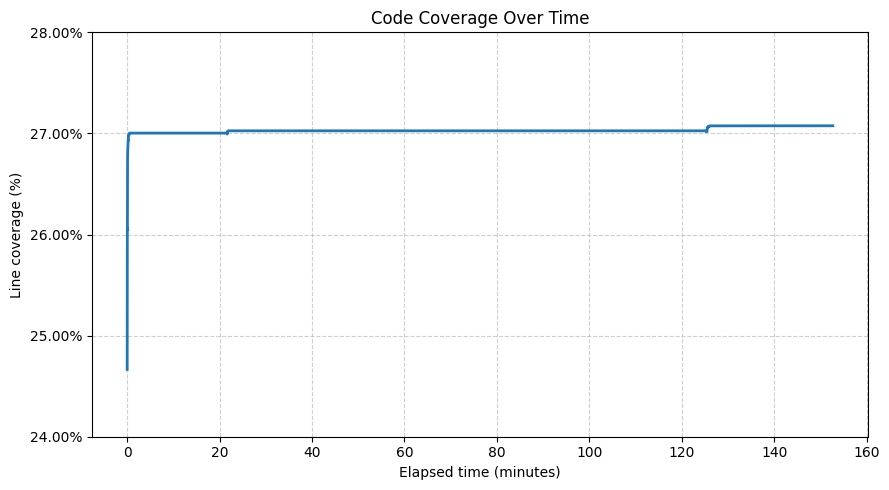}
	\caption{Code coverage over time.}
	\label{fig:code_coverage}
\end{figure}

\subsection{Code coverage over time}
Figure~\ref{fig:code_coverage} shows \wufu{} code coverage over time when fuzzing the Petstore API in white box mode.
The figure shows that the initial coverage is already high (around $\sim25\%$). This means that the fuzzer already achieves high coverage during the first runs.
Moreover, during the first minutes, the coverage reaches  $\sim27\%$, after which it remains stable for the rest of the experiment (over two hours of execution).
This means that the fuzzer immediately explores most of the code paths but struggles to find other interesting requests later.

\section{\uppercase{Limitations and future work}}
The use of \wufu{} has increased interest in other desired functionalities and experimental results, which could be obtained in future work.

Currently, \wufu{} reports the sequence of requests and responses that eventually resulted in unexpected behaviour of the SUT. In order to make analysis easier, automated deriving of a minimal sequence of requests that leads to the same behaviour is of interest. One way to do this has been described by the makers of RESTler \cite{atlidakis2019restler}. Similarly, minimising parameters to only include parts that cause unexpected behaviour is also helpful for an analyst. 

For further analysis of \wufu{}, experiments in a white-box setting can improve insights in the effectiveness of the fuzzer compared to others.

\section{\uppercase{Conclusion}}
\label{sec:conclusion}
This paper addresses some of the challenges when fuzzing in a REST API context, and the corresponding solutions provided by \wufu{}.
In order to remove the manual efforts usually needed when creating a fuzzing harness an automated harness generator was created that makes intelligent use of OpenAPI specifications.
The corpus generator generates sequences of related requests, also based on the OpenAPI specification.
Moreover, the fuzzer is able to generate interesting sequences of requests and mutate single requests and sequences of requests.

We tested our approach with different schedulers, showing a different performance in a white-box and black-box setting. After a short time of fuzzing, the response coverage becomes stable. In the white-box approach, the different responses are triggered somewhat quicker than in the black-box approach.

The fuzzer was devised with usability in mind.
Different kinds of reporting and collection of inputs sent to the fuzzer provide a vast amount of information to help with interpreting the results.
With a range of configuration options, as well as corresponding documentation, the fuzzer is ready for use.
Still, there is ample room for new developments.
We invite everyone to contribute to this fuzzer.

\section*{\uppercase{Acknowledgements}}
The publication of \wufu{} would not have been possible without the contributions of Stefan van den Berg, Alexandra Garban, Victor Li, Jacinta Moons, Jelle Nauta and Bert Jan te Paske.

\bibliographystyle{apalike}
{\small
\bibliography{references}}

\section*{\uppercase{Appendix}}\label{app:mutators}
In Table~\ref{tab:libaflmut}, the 22 used mutators from LibAFL are listed. For an explanation of their functionality, the reader is referred to the LibAFL documentation on Mutators \cite{libafl2024trait}.

\begin{table*}
    \centering
	\caption{Used mutators from LibAFL}
	\label{tab:libaflmut}
	\begin{tabular}{ll}
        \hline
		BitFlipMutator & BytesInsertCopyMutator\\
		ByteAddMutator & BytesInsertMutator\\
		ByteDecMutator & BytesRandInsertMutator\\
		ByteFlipMutator & BytesRandSetMutator\\
		ByteIncMutator & BytesSetMutator\\
		ByteInterestingMutator & BytesSwapMutator\\
		ByteNegMutator & DwordAddMutator\\
		ByteRandMutator & DwordInterestingMutator\\
		BytesCopyMutator & QwordAddMutator\\
		BytesDeleteMutator & WordAddMutator\\
		BytesExpandMutator & WordInterestingMutator\\
        \hline
	\end{tabular}
\end{table*}

In Table~\ref{tab:wufumut}, the nine additional \wufu{} mutators are listed, along with an explanation of their functionality. Apart from StringInterestingMutator, they all mutate sequences of requests.

\begin{table*}
\centering
	\caption{Used mutators from \wufu{}}
	\label{tab:wufumut}
	\begin{tabular}{lp{0.7\linewidth}l}
		\hline
		Name & Function description \\
        \hline
		AddRequestMutator & Append request to random endpoint from OpenAPI specification sequence, while filling parameters with random bytes.\\
		BreakLinkMutator & Replaces parameter value that is saved between requests and responses by a random value.\\
		DifferentMethodMutator & Changes HTTP method in existing request in sequence to another one specified in the OpenAPI specification or configured by the user.\\
		DifferentPathMutator & Changes request in sequence to use another endpoint and method from the OpenAPI specification.\\
		DuplicateRequestMutator & Duplicates request in sequence and places duplication right after original.\\
		EstablishLinkMutator & Checks for links in requests by checking whether input and return parameter names coincide and creates link when it was not already present.\\
		RemoveRequestMutator & Removes request in sequence, if sequence contains more than one request.\\
		StringInterestingMutator & Mutates string parameter value by inserting certain special values that are designed to trigger certain classes of bugs.\\
		SwapRequestsMutator & Swaps two requests in sequence.\\
        \hline
	\end{tabular}
\end{table*}

\end{document}